\documentclass{arxivpreprint}

\usepackage{amsmath,amssymb}
\usepackage{graphicx}
\usepackage[T1]{fontenc}
\usepackage[utf8]{inputenc}
\usepackage[protrusion=true,expansion=false]{microtype} % reduce overfull lines
\usepackage{booktabs}
\usepackage{cleveref}
\usepackage{siunitx}
\usepackage[section]{placeins}

\graphicspath{{figures/}}
\begin{document}

\title{An Angular Spectrum Method for Nonlinear Propagation in
Heterogeneous Tissue with Immersed Sources for Ultrasound}

\author{Gianmarco F Pinton$^{1,*}$\orcid{0000-0002-4896-1439}}

\affil{$^1$Lampe Joint Department of Biomedical Engineering, University of North Carolina at Chapel Hill and North Carolina State University, Chapel Hill, NC, USA}

\affil{$^*$Author to whom any correspondence should be addressed.}

\email{gia@email.unc.edu}


% ------------------------------------------------------------------
\begin{abstract}
A modified angular spectrum method (ASM) is developed for
three-dimensional nonlinear acoustic propagation through
heterogeneous tissue, targeting transcranial and therapeutic
ultrasound applications that include dosimetry, beam
characterization and design, and harmonic and superharmonic
analysis.  Three developments are introduced.  First,
a consistent obliquity correction is applied to both the linear and
nonlinear operators of the split-step update.  The attenuation and
dispersion filter carries a per-mode $k/k_z$ factor on the full
$(k_x, k_y, \omega)$ grid, so that every plane-wave component
accumulates absorption and phase over its true path length
$\Delta z/\cos\theta$.  The Burgers coefficient is scaled at each
step by the power-weighted mean obliquity $\langle k/k_z \rangle$ of
the current plane-wave spectrum, reducing to unity for axially
coherent beams and increasing for deeply curved fronts.  Second, the
retarded-time Burgers update is discretized with a second-order
Kurganov--Tadmor central-upwind flux using MUSCL reconstruction and
SSP-RK2 time integration, composed with the diffraction and
attenuation operators through Strang splitting with adaptive
CFL sub-cycling.  This allows stable resolution of fully developed
shocks without the temporal refinement imposed by the CFL coupling
intrinsic to FDTD.  Third, a
plane-by-plane source-injection scheme decomposes deeply curved bowl
transducers into axial slices that are injected at their correct
propagation depths, preserving the aperture-dependent
shock-formation distance that governs near-focal nonlinear
accumulation in clinical devices such as the TIPS annular phased
array.  Phase-and-amplitude screens derived from skull CT data
model transcranial aberration and insertion loss.  Three enhanced
absorbing-boundary treatments (a Wendland $C^2$ taper,
frequency-weighted spatial damping, and a first-order
Engquist--Majda super-absorbing correction) reduce boundary
reflections by a factor of 2.4 relative to a standard quadratic
absorber.  Validation against analytical solutions reproduces the
baffled-piston far-field pattern with an RMS error of 0.014 and the
focused-piston focal pressure to within 2.3\%.  In a transcranial
benchmark through an \emph{ex~vivo} human skull, the ASM matches
the Fullwave~2 focal-plane intensity with a 1.1\% RMS difference and
predicts a through-skull insertion loss of \SI{5.4}{\deci\bel}.
For a bowl transducer ($R = \SI{80}{\milli\meter}$,
$f_0 = \SI{1}{\mega\hertz}$), the plane-by-plane ASM matches the
Fullwave~2 focal depth to within 2.3\% with $9\times$ less
memory and $2.9\times$ less wall time.  The ASM is thus a practical
tool for transcranial treatment planning,
aberration correction, acoustic radiation force estimation, thermal
dose prediction, and ultrasonic neuromodulation.
\end{abstract}

% ------------------------------------------------------------------
\section{Introduction}
\label{sec:intro}

Transcranial focused ultrasound requires propagation models that
capture nonlinear distortion, frequency-dependent attenuation, and
wavefront aberration through the skull, yet remain fast enough
for treatment planning and aberration correction.  Full-wave
finite-difference time-domain (FDTD) solvers
\cite{pinton2009,pinton2007} provide high fidelity but demand large
3-D grids and long run times.  The angular spectrum method (ASM)
\cite{christopher1991,du1985,zemp2003,dagrau2011,luo2020} is an
alternative. By decomposing the field into plane-wave
components and applying the exact free-space propagator in wavenumber
space via the FFT, it avoids the paraxial approximation and
remains computationally efficient.  A further advantage of the ASM for
therapeutic applications is that it supports a fine temporal
grid independently of the spatial grid.  In FDTD methods the temporal
and spatial sampling are coupled through the CFL condition, so
resolving shocked waveforms with steep temporal gradients requires
refining the entire 3-D spatial grid.  In the ASM the retarded-time
Burgers update operates on a 1-D temporal axis at each spatial point,
so the temporal sampling can be made arbitrarily fine without
increasing the spatial grid size or the cost of the diffraction step.
This makes the ASM well suited to nonlinear propagation
in the shock-forming regime, including cubic nonlinearities in soft
solids where shear shock waves develop over sub-wavelength distances
\cite{pinton2010,giammarinaro2016,pinton2026shear}.  The principal limitation of the
ASM is that it propagates only forward and therefore cannot model
reflections, mode conversion, or multiple scattering at acoustic
interfaces.

Extensions of the ASM for therapy-planning applications have been
pursued along three parallel lines.  The first line has
addressed propagation through heterogeneous tissue.  Hybrid
angular-spectrum formulations decompose the propagation volume into
thin homogeneous slabs and insert phase-and-amplitude screens between
split-step propagations to predict linear transcranial fields
from CT-derived skull maps \cite{vyas2012}.  This
framework has been extended to hemispherical HIFU transducers, whose
curvature depth exceeds the wavelength by more than an order of
magnitude, by representing the bowl as an equivalent-source-plane
projection.  That extension established that deeply curved
transducers are tractable within a hybrid-ASM framework under
linear-propagation assumptions \cite{almquist2015}.
The second line has addressed nonlinear propagation in heterogeneous
media.  Wave-vector-frequency-domain and mixed-domain methods
\cite{jing2011,gu2018} combine ASM-style diffraction with a
Westervelt-type nonlinear operator and permit finite-amplitude
transcranial simulations in the pre-shock regime.  A related body of
nonlinear-ASM work in homogeneous media \cite{christopher1991,
du1985,zemp2003,dagrau2011} developed the retarded-time split-step
framework that underlies these extensions, and some
implementations apply an obliquity-corrected absorption term in the
plane-wave domain \cite{dagrau2011}.  The third line has addressed
curved-source field generation. Rapid continuous-wave calculations
from arbitrary curved sources \cite{treeby2018} and fast-nearfield
angular-spectrum hybrids \cite{zeng2008} provide accurate
source-plane descriptions for planning tools in the linear regime.

Despite this body of work, three limitations constrain existing ASM
implementations for strongly nonlinear transcranial or therapeutic
propagation from deeply curved bowl transducers.  First, attenuation and nonlinearity
in most nonlinear-ASM implementations are applied as if each
plane-wave component traversed the axial distance $\Delta z$ per
step rather than its physical path length $\Delta z / \cos\theta$.
An obliquity-corrected absorption filter appears in isolated
implementations \cite{dagrau2011}, but no corresponding correction
on the nonlinear operator has been reported in the heterogeneous-ASM
setting.  The resulting under-attenuation and underestimation of
nonlinear steepening for oblique components become a leading-order
error for bowls with rim-ray angles approaching $30^\circ$.  Second,
the nonlinear operator in existing heterogeneous-ASM implementations
\cite{jing2011,gu2018} uses a Westervelt-type weak-nonlinearity
update without flux-conservative shock capturing.  Strongly shocked
waveforms with steep temporal gradients, characteristic of
high-intensity transcranial focusing at multi-megapascal focal
pressures, cannot be resolved stably in that framework without
non-physical oscillation or substantial temporal refinement.  Third,
source models for deeply curved bowls in prior ASM work either
assume linear propagation \cite{almquist2015,treeby2018} or collapse
the curved surface onto a single equivalent source plane.  In the
linear regime the collapse is rigorous through Rayleigh--Sommerfeld
reciprocity.  In the nonlinear regime it discards the
aperture-dependent shock-formation distance that governs near-focal
steepening, since different rays from the aperture reach shock
formation at different axial depths.

The present work addresses these three limitations within a single
split-step solver.  The first contribution is a consistent treatment
of off-axis path length in both the linear and nonlinear operators
of the split-step update.  The attenuation and dispersion filter is
applied on the full $(k_x, k_y, \omega)$ grid with a per-mode factor
$k/k_z = 1/\cos\theta$, so that each plane-wave component
accumulates absorption and phase over its true path length
$\Delta z/\cos\theta$ rather than over the axial step $\Delta z$.
The correction shares the FFT pair used for diffraction and
therefore carries negligible additional cost.  Because the nonlinear
operator is quadratic and couples plane-wave modes, a per-mode
correction is not available.  The Burgers coefficient
$N = \beta/(2\rho_0 c_0^3)$ is instead scaled at each step by the
power-weighted mean obliquity $\langle k/k_z \rangle$ of the current
plane-wave spectrum, reducing to unity for an axially coherent beam
and increasing for strongly curved fronts.  These
corrections mitigate the leading-order path-length bias that affects
strongly focused beams and integrate with the split-step
operator composition.

The second contribution is a flux-conservative, shock-capturing
discretization of the retarded-time Burgers operator compatible with
heterogeneous ASM propagation.  The nonlinear update is implemented
with a second-order Kurganov--Tadmor central-upwind flux
\cite{kurganov2000}, using MUSCL reconstruction \cite{vanleer1979}
with a monotonized-central slope limiter and strong-stability-preserving
Runge--Kutta (SSP-RK2) time integration \cite{gottlieb2001}.  The
nonlinear step is composed with the diffraction and attenuation
operators through Strang splitting \cite{strang1968,yoshida1990}
with adaptive Courant--Friedrichs--Lewy sub-cycling.  Relative to
the Westervelt-type weak-nonlinearity updates used in prior
heterogeneous-ASM implementations, the present discretization
resolves fully developed shocks without the non-physical oscillation
that attends non-monotone schemes and without the severe temporal
refinement imposed by the CFL coupling intrinsic to FDTD.  A
first-order Rusanov flux \cite{rusanov1962} is retained as a simpler
alternative for weakly nonlinear problems.

The third contribution is a plane-by-plane source-injection scheme
for deeply curved bowl transducers.  The curved transducer surface
is decomposed into a sequence of axial slices, and each slice is
injected at its own propagation depth rather than being collapsed
onto a single equivalent source plane.  The scheme preserves the
aperture-dependent shock-formation distance that governs near-focal
nonlinear accumulation in deeply curved bowls, since rays
originating at the rim of the aperture traverse a longer path
before reaching the focus than rays originating near the axis.  The
scheme is demonstrated on the TIPS annular phased array
($R = \SI{80}{\milli\meter}$, curvature depth several wavelengths)
and compared against a three-dimensional Fullwave~2 reference
\cite{pinton2009,pinton2007}.

Phase-and-amplitude screens derived from skull CT data are inserted
between propagation steps to model transcranial aberration and
insertion loss.  Three enhanced absorbing-boundary treatments are
also introduced to suppress the periodic-FFT wrap-around artifacts
intrinsic to the ASM, namely a Wendland $C^2$ taper, a frequency-weighted
spatial damping layer, and a first-order Engquist--Majda
super-absorbing correction.  An adaptive $k$-space cosine filter
maintains stability at large axial steps.  These components are
deployed in the transcranial and bowl-transducer benchmarks where
they materially affect accuracy.  The solver is validated through
analytical benchmarks, a transcranial comparison against Fullwave~2
through an \emph{ex~vivo} human skull, and a bowl-transducer
comparison against a three-dimensional Fullwave~2 simulation of a clinical
annular phased array.

The same flux-conservative framework is extended in \Cref{app:cubic}
to the cubic Burgers equation that governs the soft-tissue
shear-shock regime \cite{pinton2010,giammarinaro2016}.  In soft
tissue the low shear-wave speed drives shock formation over
sub-wavelength distances and produces the odd-harmonic generation
characteristic of this regime
\cite{pinton2010,giammarinaro2016,tripathi2019}.  Unlike the
quadratic Burgers term of longitudinal acoustics, the cubic term
steepens both the leading and trailing edges of the waveform,
producing compound waves that require a flux-conservative
shock-capturing discretization to avoid spurious oscillation.  A
sonic-detector switch on the Kurganov--Tadmor flux retains
second-order accuracy on smooth data and resolves these compound
waves entropically.

% ------------------------------------------------------------------
\section{Methods}
\label{sec:methods}

\subsection{Split-step propagation}

The modified angular spectrum approach decomposes the propagation of a
pulsed acoustic field $p(x,y,t;z)$ into three operators applied at
each axial step $\Delta z$, namely diffraction, nonlinear distortion,
and attenuation.

In the Fourier domain $(k_x, k_y, k_t)$, the free-space diffraction
propagator is
\begin{equation}
  \hat{H}(k_x, k_y, k_t) =
  \begin{cases}
    e^{i\Delta z\bigl(k + i\sqrt{k_\perp^2 - k^2}\bigr)}, & k_\perp^2 > k^2 \\[4pt]
    e^{i\Delta z\bigl(k - \sqrt{k^2 - k_\perp^2}\bigr)},   & k_\perp^2 \le k^2
  \end{cases}
  \label{eq:propagator}
\end{equation}
where $k = \omega/c_0$ is the temporal wavenumber, $k_\perp^2 = k_x^2
+ k_y^2$, and evanescent components ($k_\perp > k$) decay
exponentially.  A cosine taper is applied near the spatial Nyquist
boundary $k_N = \pi/\Delta x$ to suppress aliasing at coarse
resolution.

The cumulative nonlinear effect is modeled as the retarded-time
Burgers equation
\begin{equation}
  \frac{\partial p}{\partial z} = -\frac{\beta}{2\rho_0 c_0^3}
  \frac{\partial p^2}{\partial t},
  \label{eq:burgers}
\end{equation}
discretized via either the first-order Rusanov (local Lax--Friedrichs)
flux \cite{rusanov1962},
\begin{equation}
  F_{j+1/2} = -\tfrac{1}{2}(p_j^2 + p_{j+1}^2)
  - \lambda_{j+1/2}(p_{j+1} - p_j),
  \label{eq:rusanov}
\end{equation}
with $\lambda_{j+1/2} = \max(|p_j|, |p_{j+1}|)$, or the second-order
Kurganov--Tadmor (KT) central-upwind scheme \cite{kurganov2000},
\begin{equation}
  F_{j+1/2}^{\mathrm{KT}} =
  \frac{a^+ f(p_j) - a^- f(p_{j+1})
        - a^+ a^-(p_{j+1} - p_j)}{a^+ - a^-},
  \label{eq:kt}
\end{equation}
where $a^{\pm}$ are the one-sided local wave speeds.  To achieve
second-order spatial accuracy, the cell-edge values in \cref{eq:kt}
are obtained from a piecewise-linear MUSCL reconstruction
\cite{vanleer1979} with an MC (monotonized central) limiter applied to
each cell.  The reconstructed left and right states
$p_{j+1/2}^L = p_j + \tfrac{1}{2}\,\sigma_j$ and
$p_{j+1/2}^R = p_{j+1} - \tfrac{1}{2}\,\sigma_{j+1}$ replace the
cell-center values in the flux formula, where $\sigma_j$ is the
limited slope.  The resulting semi-discrete system is advanced in $z$
with a two-stage strong-stability-preserving Runge--Kutta method
(SSP-RK2) \cite{gottlieb2001},
\begin{align}
  p^{(1)} &= p^{(n)} + \Delta z\,\mathcal{L}(p^{(n)}), \notag \\
  p^{(n+1)} &= \tfrac{1}{2}\,p^{(n)} + \tfrac{1}{2}\bigl[
    p^{(1)} + \Delta z\,\mathcal{L}(p^{(1)})\bigr],
  \label{eq:ssprk2}
\end{align}
where $\mathcal{L}$ denotes the spatial flux divergence.  The
combination of MUSCL reconstruction and SSP-RK2 gives second-order
accuracy for the isolated nonlinear problem in smooth regions.

The retarded-time Burgers update acts pointwise in the transverse
$(x, y)$ plane and implicitly treats $\Delta z$ as the physical path
length over which nonlinear steepening accumulates.  For a deeply
curved source, plane-wave components propagate at angles $\theta$
from the $z$ axis and their true per-step path length is
$\Delta z / \cos\theta$, so the pointwise update underestimates the
nonlinear accumulation of oblique components.  Unlike attenuation,
nonlinearity mixes plane-wave modes and cannot be corrected with a
per-mode factor.  A beam-averaged correction is therefore applied at
every step, in which the Burgers coefficient
$N = \beta/(2\rho_0 c_0^3)$ is scaled by the power-weighted mean
obliquity
\begin{equation}
  \bigl\langle k / k_z \bigr\rangle =
  \frac{\displaystyle\sum_{k_x, k_y, \omega} |\tilde p(k_x, k_y, \omega)|^2\,
        k(\omega)/k_z(k_x, k_y, \omega)}
       {\displaystyle\sum_{k_x, k_y, \omega} |\tilde p(k_x, k_y, \omega)|^2},
  \label{eq:obliquity_mean}
\end{equation}
evaluated on the field entering each nonlinear sub-step, with the
sum restricted to propagating modes.  The scaling
$N_\mathrm{eff} = N\,\langle k/k_z\rangle$ reduces to unity for an
axially coherent beam and increases the effective nonlinear path
length for strongly curved fronts.

Attenuation and dispersion are applied as a plane-wave filter derived
from the power-law absorption coefficient $\alpha(f) = \alpha_0 f^n$
with Kramers--Kronig-consistent dispersion \cite{pinton2009}.  A
frequency-only filter $\exp[-(\alpha + i\alpha^*)\,\Delta z]$ assumes
every plane-wave component travels the axial distance $\Delta z$ per
step, but a component propagating at angle $\theta$ from the $z$
axis traverses $\Delta s = \Delta z / \cos\theta = \Delta z\,k/k_z$,
where $k = \omega/c_0$ and $k_z = \sqrt{k^2 - k_\perp^2}$.  For
deeply curved sources this obliquity is significant.  Without
correction, the outer rays of a bowl with a rim-ray angle of
$30^\circ$ are under-attenuated by a factor of
$1/\cos 30^\circ \approx 1.15$ per step.  The filter is therefore built on the full
$(k_x, k_y, \omega)$ grid and applied in the same $(k_x, k_y)$
domain as the diffraction propagator,
\begin{equation}
  \hat{A}(k_x, k_y, \omega) =
  \exp\!\Bigl[-(\alpha + i\alpha^*)\,\Delta z\,\frac{k}{k_z}\Bigr],
  \label{eq:attenuation}
\end{equation}
so that each plane-wave component sees attenuation and dispersion
over its true path length.  Evanescent components ($k_\perp^2 > k^2$)
are left unchanged since the diffraction propagator already decays
them.

These three operators are composed via Strang splitting
\cite{strang1968,leveque2002} as
\begin{equation}
  p^{(n+1)} = \mathcal{A}_{1/2}\,\mathcal{H}_{1/2}\,
  \mathcal{N}\,
  \mathcal{H}_{1/2}\,\mathcal{A}_{1/2}\;p^{(n)},
  \label{eq:strang}
\end{equation}
which achieves second-order accuracy in $\Delta z$
\cite{yoshida1990}.

\subsection{Absorbing boundary layer}

The ASM uses global FFTs that impose periodic boundary conditions, so
energy reaching the domain edge wraps to the opposite side.  The
standard mitigation is a multiplicative absorbing boundary layer (ABL)
that damps the field near the edges.  Three improvements are
introduced.

First, the quadratic taper profile $1 - s^2$ is replaced with the
Wendland $C^2$ function \cite{wendland1995},
\begin{equation}
  w(s) = 1 - 10s^3 + 15s^4 - 6s^5,
  \label{eq:wendland}
\end{equation}
which has continuous first and second derivatives at the taper onset
($s = 0$), eliminating the weak impedance discontinuity that generates
spurious reflections in the quadratic profile.

Second, the spatial damping is made frequency-dependent by raising the
profile to a frequency-dependent exponent,
\begin{equation}
  \mathrm{ABL}_f(x,y) = \bigl[\mathrm{ABL}(x,y)\bigr]^{f_0/\max(f,\,f_{\min})},
  \label{eq:freq_abl}
\end{equation}
which strengthens the damping at low frequencies where the boundary
layer spans fewer wavelengths.

Third, a first-order Engquist--Majda \cite{engquist1977} directional
decomposition is applied in the boundary region to selectively remove
incoming wave components.  At each temporal frequency $\omega$, the
incoming pressure is estimated as
\begin{equation}
  P_{\mathrm{in}} = \tfrac{1}{2}\Bigl(P + \frac{c_0}{i\omega}
  \frac{\partial P}{\partial x}\Bigr),
  \label{eq:engquist}
\end{equation}
and the correction $P \leftarrow P - \alpha_s\,w_{\mathrm{bdy}}\,
P_{\mathrm{in}}$ is applied in the boundary layer, where
$\alpha_s \in [0,1]$ controls the strength of the incoming-wave
removal.

\subsection{Heterogeneous propagation}

Spatially varying sound speed and attenuation are modeled using thin
phase-and-amplitude screens inserted at each propagation step.  Each
screen represents a tissue layer of thickness $d$ with maps $c(x,y)$
and $\rho(x,y)$ derived from imaging data.  The phase shift at each
temporal frequency is
\begin{equation}
  \Delta\varphi(x,y,f) = 2\pi\,d\,
  \biggl(\frac{1}{c(x,y)} - \frac{1}{c_0}\biggr) \cdot f,
  \label{eq:phase_screen}
\end{equation}
which is linear in frequency, corresponding to a
frequency-independent group delay through each screen.
Each screen also carries a real-valued amplitude factor
$A(x,y) \in (0,1]$ that accounts for attenuation and
impedance-mismatch losses,
\begin{equation}
  A(x,y) = \exp\!\bigl[-\alpha_{\mathrm{Np}}(x,y)\,d\bigr]
  \;\cdot\;
  \frac{2\,Z(x,y)}{Z(x,y) + Z_0},
  \label{eq:amplitude_screen}
\end{equation}
where $\alpha_{\mathrm{Np}} = \alpha\,f_{\mathrm{MHz}} \times
\left(\tfrac{\ln 10}{20}\times 100\right)$ converts the local attenuation
$\alpha(x,y) = \alpha_{\mathrm{tissue}} + f_{\mathrm{bone}}(x,y)\,
(\alpha_{\mathrm{bone}} - \alpha_{\mathrm{tissue}})$ from
\si{\dB\per\cm\per\MHz} to \si{Np\per\m}, $f_{\mathrm{bone}}$ is the
bone volume fraction, and $Z = \rho\,c$ is the acoustic impedance.
Frequency-dependent amplitude is applied as $A^{f/f_0}$ so that higher
harmonics experience proportionally greater loss.

The field is transformed to the temporal frequency domain via a
real-valued Fourier transform, multiplied by
$A^{f/f_0}\,e^{i\Delta\varphi}$, and transformed back.  Multiple screens can be placed at arbitrary depths
to model extended heterogeneous paths via the split-step Fourier
approach \cite{martin1988}.  In the transcranial benchmark below, the
screens are derived directly from skull CT slices rather than from a
random-field model.

\subsection{Intensity-loss tracking}

In therapeutic ultrasound, the absorbed acoustic energy drives both
radiation force and thermal bioeffects.  The standard plane-wave
approximations for the volume rate of heat deposition,
$q = 2\alpha I$ \cite{nyborg1981}, and the radiation force body-force
density, $q_{\mathrm{ARF}} = 2\alpha I / c$, assume a single
attenuation coefficient $\alpha$ evaluated at the fundamental
frequency.  In the nonlinear regime, however, cumulative harmonic
generation transfers energy to higher frequencies where the
power-law attenuation $\alpha(f) = \alpha_0 f^n$ is substantially
larger.  The local frequency content of the wave therefore changes
with propagation distance, and the effective attenuation can no
longer be represented by a single $\alpha$.

To avoid this approximation, the solver tracks the spatially
resolved intensity drop across the attenuation operator at each
propagation step.  With Strang splitting the attenuation operator
is applied as two half steps that bracket the diffraction and
nonlinear sub-steps.  Let $I_{0}$, $I_{1}$, $I_{2}$, and $I_{3}$
denote the per-pixel time-integrated intensity $\sum_{t} p^{2}$ at
the four breakpoints, where $I_{0}\!\to\!I_{1}$ and
$I_{2}\!\to\!I_{3}$ are the two attenuation half steps.  The
dissipated energy at the step is
\begin{equation}
  \Delta I_{\mathrm{loss}}(x,y;z) =
  \max\!\bigl(0,\, I_{0}-I_{1}\bigr) +
  \max\!\bigl(0,\, I_{2}-I_{3}\bigr).
  \label{eq:loss}
\end{equation}
Each half-step difference is guaranteed non-negative by Parseval's
identity and the unitarity of the dispersion phase rotation.  The
$\max(0,\,\cdot)$ guards against float32 round-off, which can
produce slightly negative values where the true loss is small.
Computing the loss across the attenuation operator alone, rather
than across the full propagation step, isolates absorption from the
diffractive lateral redistribution and shock-induced energy
transport that a step-wise total-intensity difference would
conflate with dissipation.  The resulting field is a per-pixel
measure of absorbed energy that implicitly accounts for the full
harmonic content of the wave without requiring an explicit
frequency decomposition or a single-frequency $\alpha$
assumption.  The loss field can be used directly as a heat source
for thermal simulations or as input for radiation-force
calculations.

\subsection{Implementation}
\label{sec:methods:implementation}

The solver is implemented in Python using JAX \cite{jax2018} for GPU
acceleration.  All frequency-domain operators, including the
diffraction propagator, the obliquity-corrected attenuation filter,
and the obliquity map used for the nonlinear correction, are
precomputed once and reused at each propagation step.

% ------------------------------------------------------------------
\section{Validation}
\label{sec:validation}

The solver is validated against analytical solutions for diffraction,
attenuation, and nonlinearity, followed by boundary-treatment and
convergence studies.  All simulations use $\Delta x = \lambda/5$
unless otherwise noted.

\subsection{Diffraction}
\label{sec:val_diffraction}

The baffled-piston and focused-piston benchmarks have been used
extensively to validate angular spectrum propagators
\cite{christopher1991,zemp2003,du1985} and serve here to verify the
present implementation.

\begin{figure}[htbp]
  \centering
  \includegraphics[width=\columnwidth]{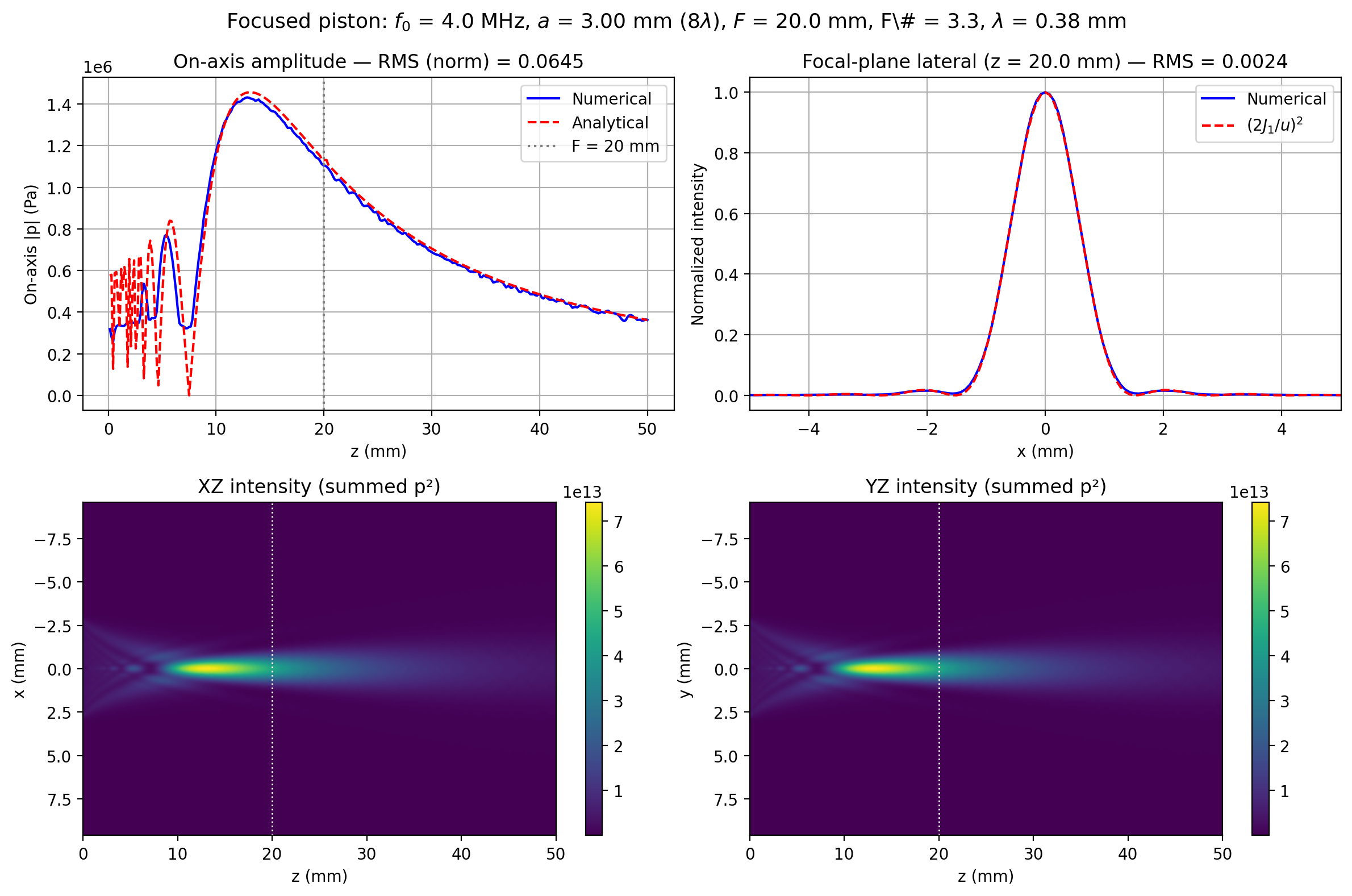}
  \caption{Focused piston diffraction ($a = 8\lambda$,
    $F = \SI{20}{\milli\meter}$, $F\#\,\approx 3.3$).  The top row
    shows the on-axis amplitude versus the paraxial solution (left,
    RMS 0.0645) and the focal-plane lateral profile versus the Airy
    pattern (right, RMS 0.002).  The bottom row shows the $x$--$z$
    and $y$--$z$ time-integrated intensity.}
  \label{fig:focused_piston}
\end{figure}

A spherically focused piston ($a = 8\lambda$,
$F = \SI{20}{\milli\meter}$, $F\#\,\approx 3.3$) at
$f_0 = \SI{4}{\MHz}$ was propagated with per-pixel geometric delays
$\tau = (F - \sqrt{F^2 + r^2})/c_0$ (\cref{fig:focused_piston}).
The numerical focal pressure matches the paraxial limit
$p_0\,\pi a^2/(\lambda F)$ to within 2.3\%, and the focal-plane
lateral profile matches the Airy pattern with an RMS error of 0.002.
An error-floor study confirmed that the residual on-axis RMS of
$\sim$0.065 reflects the mismatch between the paraxial and full-wave
solutions at this $F$-number rather than numerical error.  Refining
$\Delta z$, widening the time window, and narrowing the pulse
bandwidth all leave the error unchanged.

For the unfocused baffled piston ($a = 6\lambda$), the on-axis
amplitude was compared against the Rayleigh--Fresnel solution
\cite{kinsler2000},
\begin{equation}
  P_{\mathrm{cw}}(z) = -p_0\bigl(e^{ik_0 a^2/(2z)} - 1\bigr)\,
  e^{-ik_0 z},
  \label{eq:rayleigh}
\end{equation}
yielding a normalized RMS error of 0.086, with the numerical solution
reproducing the oscillatory Fresnel structure beyond the transition
distance $z_t \approx a^2/\lambda$.  In the far field, the lateral
intensity matches the Fraunhofer pattern
$(2J_1(k_0 a r/z) / k_0 a r/z)^2$ with RMS error 0.014.

\subsection{Attenuation and dispersion}
\label{sec:val_attenuation}

Accurate modeling of frequency-dependent attenuation is essential for
therapeutic ultrasound, where nonlinear harmonic generation
redistributes energy to higher frequencies that are absorbed more
rapidly \cite{pinton2009,treeby2012}.  The associated
Kramers--Kronig-consistent dispersion determines the local phase
velocity and must be modeled correctly to avoid cumulative waveform
distortion over clinically relevant propagation distances
\cite{waters2005,szabo1995}.

\begin{figure}[htbp]
  \centering
  \includegraphics[width=\columnwidth]{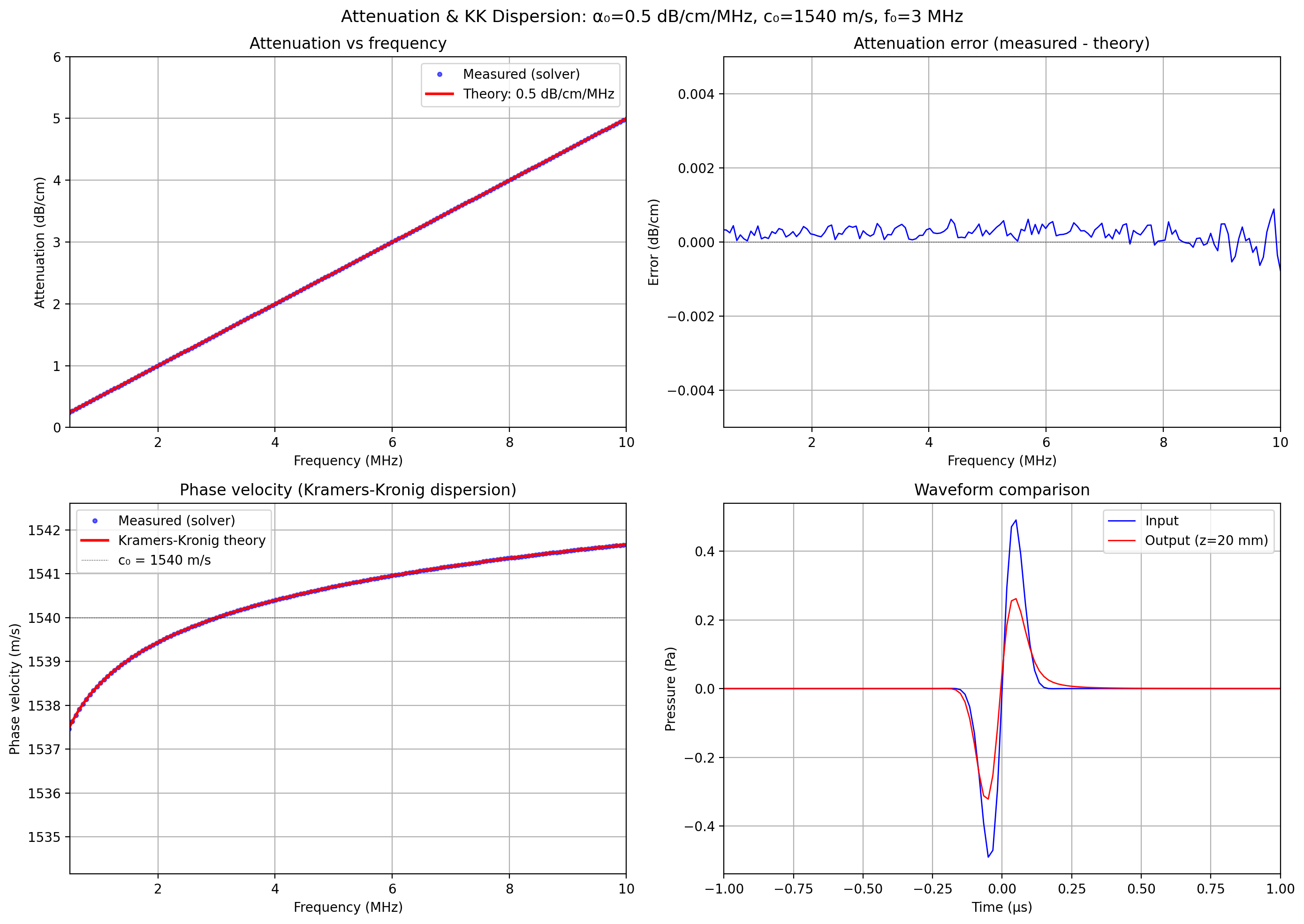}
  \caption{Attenuation and Kramers--Kronig dispersion validation
    (0.5--\SI{10}{\MHz}).  The top row shows the measured and
    theoretical attenuation and their difference.  The bottom row
    shows the phase velocity and a waveform comparison.}
  \label{fig:test10}
\end{figure}

A broadband pulse was propagated \SI{20}{\mm} with
$\alpha_0 = \SI{0.5}{\dB\per\cm\per\MHz}$ and the per-frequency
attenuation and phase velocity were measured from the input/output
spectra (\cref{fig:test10}).  The measured attenuation matches the
$\alpha_0 f$ power law with RMS error
\SI{0.0003}{\dB\per\cm} over 1--\SI{10}{\MHz}, and the phase velocity
follows the Kramers--Kronig dispersion relation across the full
0.5--\SI{10}{\MHz} range, crossing $c_0$ at $f_0$ as expected.

The $k/k_z$ obliquity correction of \cref{eq:attenuation} was
verified on representative transcranial parameters
($f = \SI{1}{\MHz}$,
$\alpha_0 = \SI{5}{\dB\per\MHz\per\cm}$, $n = 1.1$,
$\Delta z = \SI{2}{\milli\meter}$) by evaluating the filter on
individual plane-wave components of known obliquity.  The on-axis
filter magnitude is 0.898, matching the uncorrected frequency-only
filter exactly.  At $\theta = 29^\circ$, the obliquity-corrected
magnitude is 0.885 against 0.898 without correction, a ratio of
0.985 that reproduces the analytical $|\hat{A}|^{1/\cos\theta}$
scaling to round-off.  The $1.5\%$ additional attenuation per step
at $\theta = 29^\circ$ accumulates to tens of percent over a typical
transcranial beam path, justifying the per-mode treatment for
deeply curved sources.

\subsection{Nonlinearity}
\label{sec:val_nonlinearity}

The Rusanov and KT flux schemes were compared in two regimes.  In a
moderate-amplitude coupled propagation ($p_0 = \SI{30}{\kPa}$,
$\beta = 3.5$, \SI{30}{\mm}), both schemes produced nearly identical
on-axis intensity and harmonic spectra, confirming that the KT
limiter does not introduce excessive dissipation.

\begin{figure}[htbp]
  \centering
  \includegraphics[width=\columnwidth]{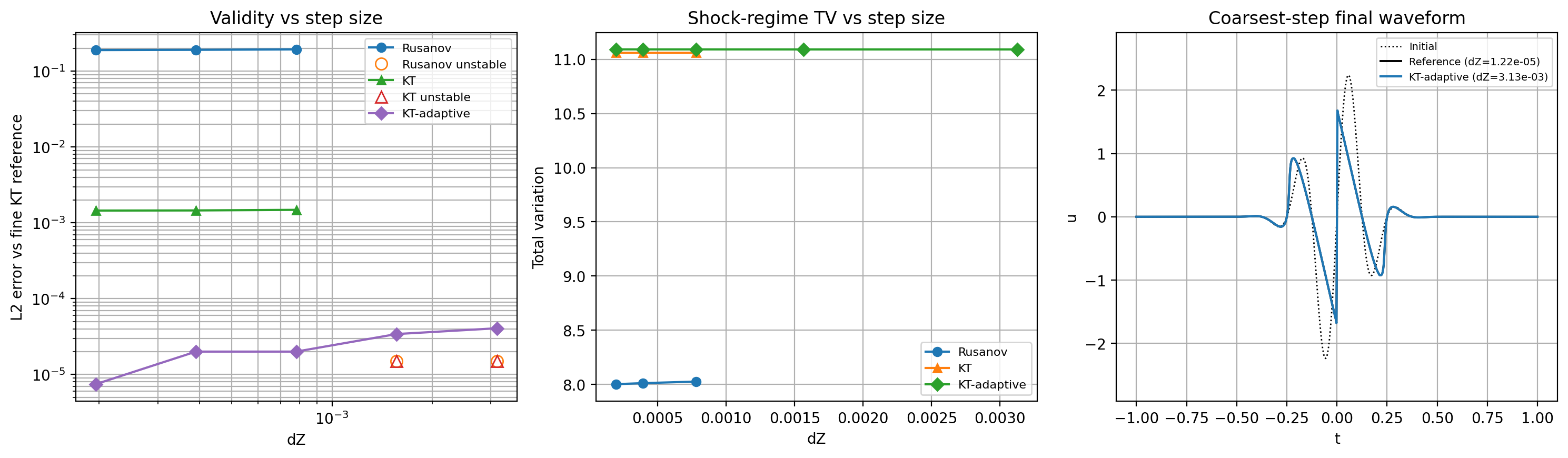}
  \caption{Shock-forming Burgers problem, showing the $L^2$ error
    versus step size (left), the total variation (center), and the
    coarsest-step waveforms (right).  The adaptive KT scheme is
    stable and accurate at all step sizes.}
  \label{fig:test3}
\end{figure}

In the shock-forming regime (\cref{fig:test3}), an isolated Burgers
step-size study compared Rusanov, KT (fixed step), and KT with
adaptive CFL sub-cycling.  Both fixed-step schemes were unstable at
coarse steps.  The adaptive KT scheme was stable at all step sizes
and the most accurate, with $L^2$ errors of $\sim 10^{-5}$ (four
orders of magnitude below Rusanov).

\begin{figure}[htbp]
  \centering
  \includegraphics[width=\columnwidth]{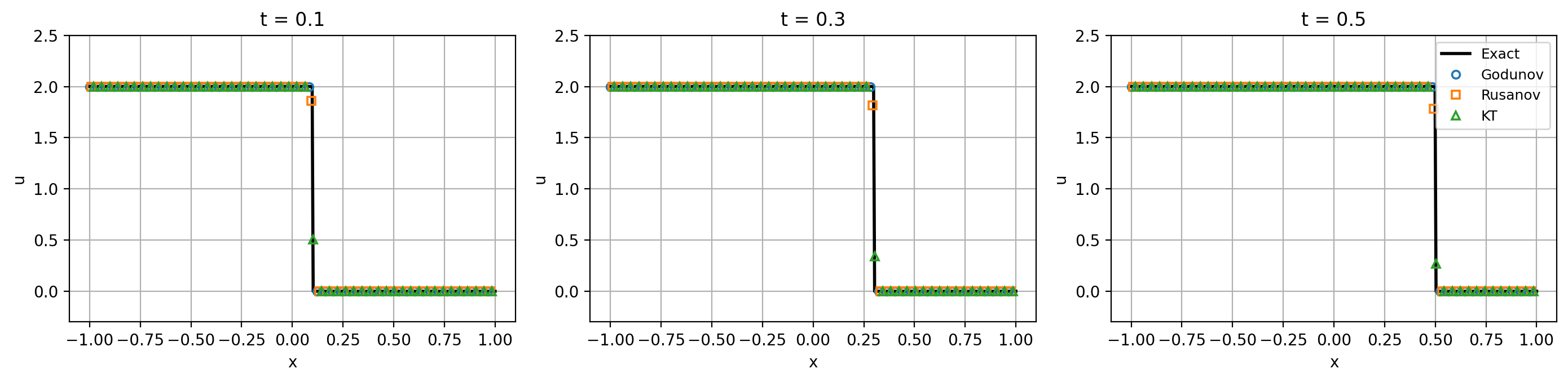}
  \caption{Riemann problem at $t = 0.1, 0.3, 0.5$.  Rusanov (circles)
    and KT (squares) overlaid on the exact solution (solid).}
  \label{fig:riemann}
\end{figure}

A separate Riemann-problem benchmark ($u_L = 2$, $u_R = 0$, 400
cells) confirmed that the KT flux resolves the shock more sharply
than Rusanov ($L^2$ error 0.036 vs.\ 0.044 at $t = 0.5$), with
both schemes capturing the correct shock speed (\cref{fig:riemann}).

The beam-averaged obliquity correction of
\cref{eq:obliquity_mean} was verified on synthetic tilted plane
waves at a range of angles.  At $\theta = 0^\circ$ the estimator
returned $\langle k/k_z\rangle = 1.001$, and at
$\theta = 15^\circ$, $30^\circ$, and $45^\circ$ it returned 1.057,
1.177, and 1.434, respectively, against the analytical values
$1/\cos\theta = 1.000$, 1.035, 1.155, and 1.414.  The small residual
at nonzero angle is consistent with spectral leakage on the
discrete wavenumber grid and decreases as the spatial FFT size
increases.  The estimator therefore recovers the correct obliquity
factor without bias on a pure plane wave and returns unity for an
axially coherent field, as required.

\subsection{Boundary treatments}
\label{sec:val_boundaries}

\begin{figure}[htbp]
  \centering
  \includegraphics[width=0.95\columnwidth]{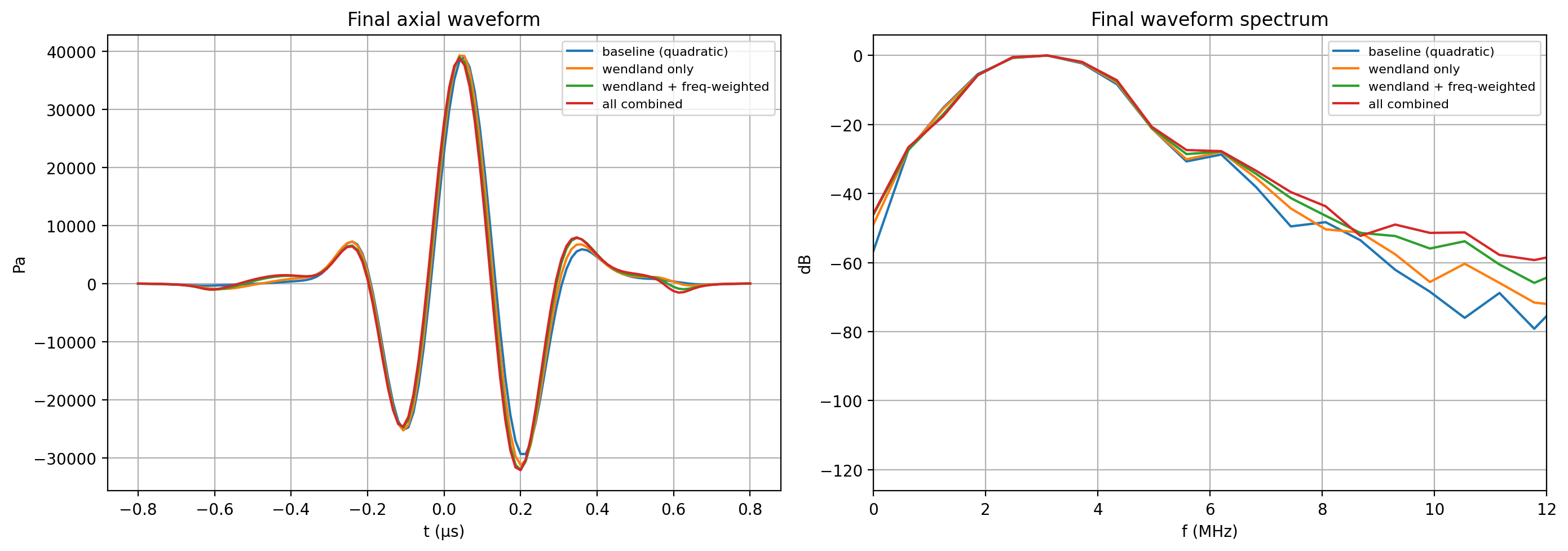}
  \caption{Combined boundary with nonlinear propagation, showing the
    final waveform (left) and spectrum (right) for four boundary
    configurations.  The interior waveform is invariant and harmonic
    levels match to $<\!1$\,dB.}
  \label{fig:bdy5_profiles}
\end{figure}

The three boundary improvements were evaluated using apertures wide
enough to drive significant energy toward the domain edges.
\Cref{tab:bdy_reflection} shows the boundary-to-center intensity
ratio as each improvement is added in turn.  The combined
configuration reduces reflections by $2.4\times$ relative to the
quadratic baseline.  A frequency-resolved comparison confirmed that
the frequency-weighted damping preferentially targets low-frequency
components (13\% reduction at \SI{1}{\MHz} vs.\ 3\% at
\SI{3}{\MHz}).  When tested with nonlinear propagation
($p_0 = \SI{30}{\kPa}$, $\beta = 3.5$), the combined boundary
recovers $\sim\!20$\% more on-axis focal intensity than the
quadratic baseline, preserves the interior waveform shape, and
matches the harmonic spectrum to within 1\,dB
(\cref{fig:bdy5_profiles}).

\begin{table}[htbp]
  \centering
  \caption{Boundary reflection ratio for four configurations.}
  \label{tab:bdy_reflection}
  \begin{tabular}{@{}lr@{}}
  \toprule
  Configuration & Reflection ratio \\
  \midrule
  Quadratic (baseline)      & $1.83 \times 10^{-2}$ \\
  Wendland $C^2$            & $1.35 \times 10^{-2}$ \\
  + Freq.\ weighted         & $0.94 \times 10^{-2}$ \\
  + Super-absorbing          & $0.75 \times 10^{-2}$ \\
  \bottomrule
  \end{tabular}
\end{table}

\subsection{Convergence}
\label{sec:val_convergence}

\begin{figure}[htbp]
  \centering
  \includegraphics[width=\columnwidth]{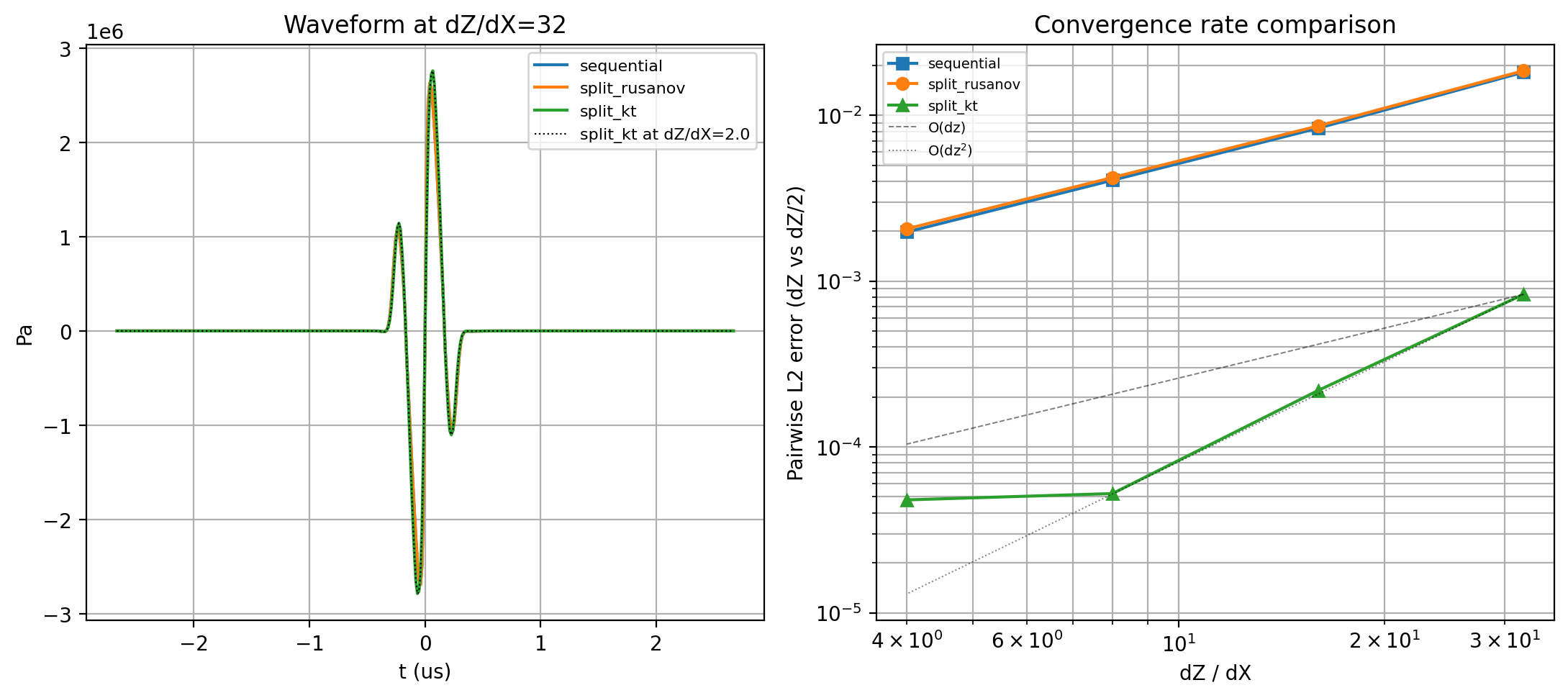}
  \caption{Step-size convergence, showing the coarsest waveforms
    (left) and the pairwise $L^2$ error (right).  Strang+KT achieves
    rate $\approx 2$.}
  \label{fig:test6}
\end{figure}

A pairwise convergence study in the strongly nonlinear regime
($p_0 = \SI{3}{\mega\pascal}$, $\beta = 3.5$, \SI{15}{\mm}
propagation) compared the $\Delta z$-vs-$\Delta z/2$ error for
sequential Rusanov, Strang+Rusanov, and Strang+KT
(\cref{fig:test6}).  Sequential and split-Rusanov converge at rate
$\approx 1$, while Strang+KT converges at rate $\approx 2$, consistent
with the expected second-order Strang splitting accuracy.

% ------------------------------------------------------------------
\section{Transcranial propagation}
\label{sec:transcranial}

To validate the phase-and-amplitude-screen extension, focused
ultrasound transmission through an \emph{ex~vivo} human skull is
simulated and compared against Fullwave~2, a validated
finite-difference time-domain solver
\cite{pinton2009,pinton2007}.

\subsection{Setup}
\label{sec:transcranial:setup}

A 1024-element sparse phased array (\SI{750}{\micro\meter} pitch) is
focused at \SI{50}{\milli\meter} depth through a temporal-bone slab
extracted from the Halle skull microCT dataset
\cite{halle_skull,mccall2023sparse,pinton2012}.  The sparse
connector geometry spans approximately
\SI{65}{\milli\meter} by \SI{65}{\milli\meter} in the transmit
plane.  The center frequency is $f_0 = \SI{1}{\mega\hertz}$, the source
pressure is $p_0 = \SI{1.5}{\mega\pascal}$, and the excitation is a
2-cycle Gaussian-windowed tone burst with geometric focusing delays.
After rasterization onto the ASM grid, the aperture occupies 9127
active source-plane grid points with a delay spread of
\SI{6.5}{\micro\second}.

The skull CT (voxel size $\approx\!\SI{0.2}{\milli\meter}$) is first
resampled into a beam-aligned slab centered at
$(140.6,118.5,85.0)$~mm in LPS coordinates and oriented with inward
normal $(-0.954,-0.298)$ and tangent $(-0.298,0.954)$ so that the beam
propagates approximately normal to the extracted bone layer.  The slab
extent is
\SI{66}{\milli\meter} by \SI{66}{\milli\meter} by \SI{65}{\milli\meter}
(lateral, elevational, depth), which at the ASM sampling of
$\Delta x = \lambda/5 \approx \SI{0.308}{\milli\meter}$ yields an
initial $214 \times 214 \times 211$ beam-aligned skull map before
lateral resampling to the propagation grid.

Hounsfield units are mapped linearly to acoustic properties using
$c = 1540 + f_{\mathrm{bone}}(2900 - 1540)$~m/s and
$\rho = 1000 + f_{\mathrm{bone}}(2200 - 1000)$~kg/m$^3$, where
$f_{\mathrm{bone}} = \mathrm{clip}[(HU - 700)/(1973 - 700),\,0,\,1]$.
Each extracted depth slice is converted into a phase-and-amplitude
screen (\cref{eq:phase_screen,eq:amplitude_screen}).  The phase term is
set by the local sound-speed contrast, while the amplitude term combines
bone attenuation
($\alpha_{\mathrm{bone}} = \SI{8}{\dB\per\cm\per\MHz}$),
soft-tissue attenuation
($\alpha_{\mathrm{tissue}} = \SI{0.5}{\dB\per\cm\per\MHz}$), and
normal-incidence impedance transmission derived from the density map.
This screen construction incorporates cumulative skull-induced
phase aberration and insertion loss without sacrificing the
efficiency of the FFT-based ASM.

The ASM grid uses
$\Delta x = \Delta y = \Delta z = \lambda/5 \approx \SI{0.308}{\milli\meter}$,
giving a $215 \times 215$ lateral grid and 212 propagated axial planes
over \SI{65}{\milli\meter}.  The temporal sampling is
$\Delta t = \SI{40}{\nano\second}$ with 501 time samples.  Propagation
uses the split-step Kurganov--Tadmor formulation together with the
Wendland $C^2$ boundary taper, frequency-weighted damping, and the
super-absorbing boundary correction.

\subsection{Fullwave~2 reference}
\label{sec:transcranial:fdtd}

The reference Fullwave~2 simulation \cite{pinton2007} uses the same
skull CT, transducer geometry, and transmit waveform, solved on a
$422 \times 429 \times 429$ grid at
$\Delta x = \SI{0.154}{\milli\meter}$ (i.e.\ $\lambda/10$) with 4221
time steps.  The source is applied over three depth layers and uses the
same 1\,MHz, 2-cycle, \SI{1.5}{\mega\pascal} transmit pulse and
geometric focusing delays as the ASM model.  The saved pressure field
is downsampled by a factor of four in space and seven in time, yielding
a $106 \times 108 \times 108$ output grid over 603 recorded time
snapshots.  For comparison with the ASM time-integrated intensity, the
Fullwave~2 metric is formed as $\sum_t p^2 \Delta t$ on the recorded series
using the effective saved sampling interval
$\Delta t_{\mathrm{FDTD}} = \SI{140}{\nano\second}$, and then
interpolated onto the ASM grid.

\subsection{Results}
\label{sec:transcranial:results}

\begin{figure}[htbp]
  \centering
  \includegraphics[width=\columnwidth]{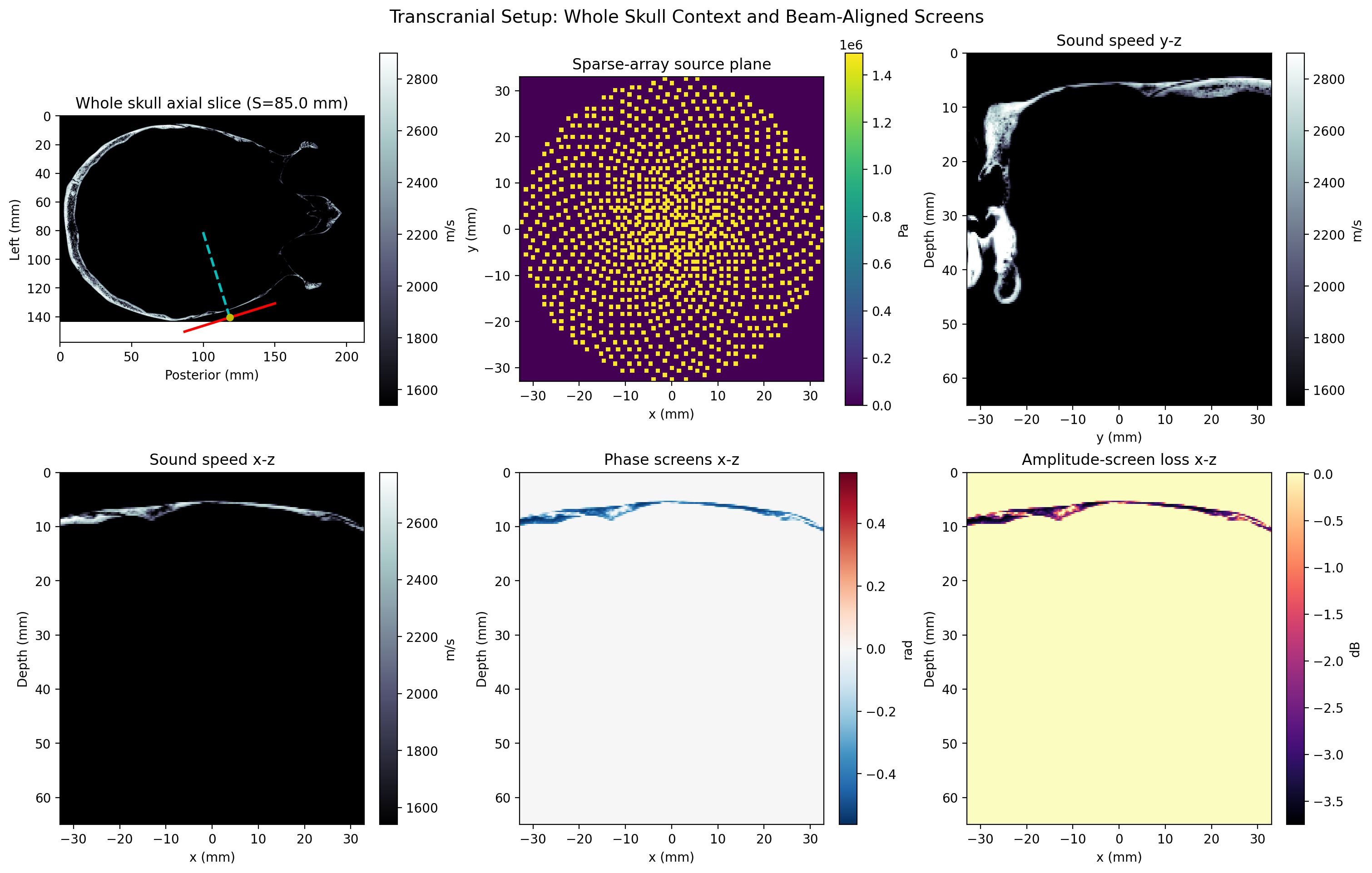}
  \caption{Transcranial simulation setup.  The top row shows the
    whole-skull axial reference slice with the beam axis overlaid
    (left), the source-plane pressure magnitude from the
    1024-element sparse array (center), and the beam-aligned
    sound-speed map in the $y$--$z$ plane (right).  The bottom row
    shows the sound-speed map in the $x$--$z$ plane (left) and the
    stacked phase-screen and amplitude-screen loss maps in the
    $x$--$z$ plane (center and right).}
  \label{fig:transcranial_setup}
\end{figure}

\begin{figure}[htbp]
  \centering
  \includegraphics[width=\columnwidth]{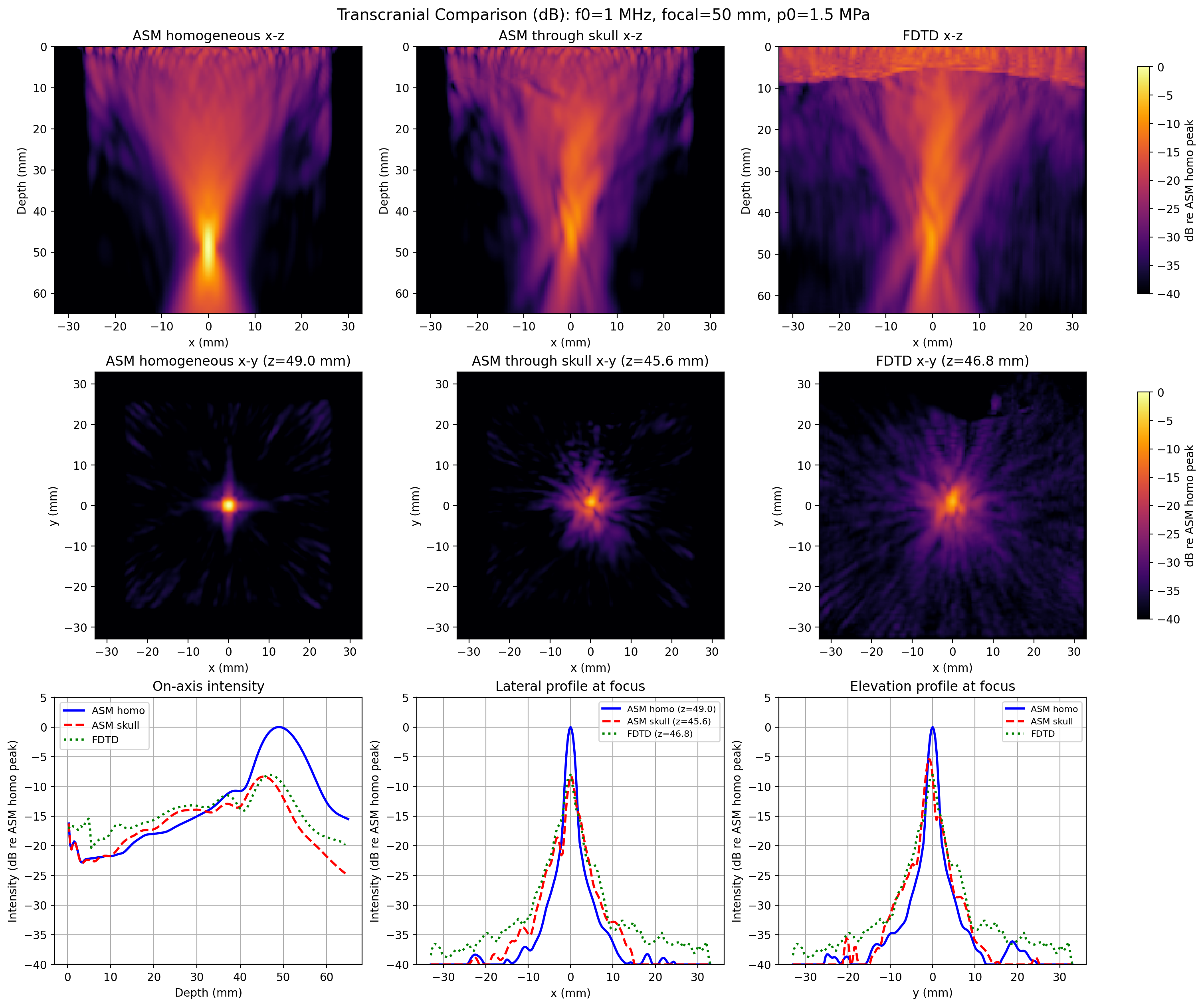}
  \caption{Transcranial comparison referenced to the homogeneous ASM
    field.  The top row shows axial ($x$--$z$) beam profiles for the
    homogeneous ASM, the through-skull ASM, and Fullwave~2.  The
    middle row shows the corresponding focal-plane ($x$--$y$)
    intensity maps at each method's focal depth.  The bottom row
    shows on-axis, lateral ($x$), and elevational ($y$) profiles in
    dB.  The Fullwave~2 intensity is weighted by its recorded
    temporal sampling interval before normalization so the dB
    offsets are consistent with the ASM time-integrated intensity.}
  \label{fig:transcranial_profiles}
\end{figure}

Figures~\ref{fig:transcranial_setup} and \ref{fig:transcranial_profiles}
summarize the transcranial setup and compare the Fullwave~2 and ASM results
on a common dB scale.
The homogeneous ASM case focuses at
$z = \SI{49.0}{\milli\meter}$, the through-skull ASM case at
$z = \SI{45.6}{\milli\meter}$, and the Fullwave~2 reference at
$z = \SI{46.8}{\milli\meter}$.  The skull therefore shifts the focus
proximally by about \SI{3.4}{\milli\meter} relative to homogeneous
propagation, while the heterogeneous ASM prediction remains within
\SI{1.2}{\milli\meter} (2.6\%) of the Fullwave~2 focal depth.

The focal-plane agreement is strong.  After weighting the Fullwave~2
intensity by its recorded temporal sampling interval so that both
solvers represent the same $\sum_t p^2 \Delta t$ quantity, the
normalized focal-plane RMS difference is 0.011 (1.1\%).  The axial
$x$--$z$ beam comparison gives an RMS difference of 0.045.  The
remaining discrepancy is concentrated in the pre-focal region and near
the skull surface, where the Fullwave~2 solution contains reflections,
interface reverberation, and mode conversion that are absent from the
forward-only ASM.

The amplitude-screen extension also captures the overall insertion
loss.  Relative to the homogeneous ASM reference, the through-skull
ASM focal peak is reduced by \SI{5.4}{\deci\bel}.  On the common
homogeneous-reference dB scale used in
\cref{fig:transcranial_profiles}, the Fullwave~2 and ASM through-skull focal
peaks differ by only about 0.3\,dB.  The predicted insertion loss
falls within the measured range of
\SIrange{5}{10}{\deci\bel}~\cite{pinton2012}, which indicates that
the heterogeneous phase-and-amplitude-screen extension captures the
first-order effects of skull aberration and attenuation and
retains the computational efficiency of the ASM.

% ------------------------------------------------------------------
\section{Immersed bowl transducer model}
\label{sec:bowl}

\subsection{Motivation}
\label{sec:bowl:motivation}

The sparse-array validation of \cref{sec:transcranial} uses a
geometrically flat source aperture. All elements lie in a single
$z = 0$ plane, and focusing is achieved entirely through electronic
time delays.  Therapeutic transducers such as the TIPS (Therapeutic
Imaging Probe and Sonication, Philips) employ a spherical bowl
geometry with a radius of curvature $R = \SI{80}{\milli\meter}$, an
annular aperture ($r_{\mathrm{in}} = \SI{20.5}{\milli\meter}$,
$r_{\mathrm{out}} = \SI{46}{\milli\meter}$), and eight concentric
annular elements.  The bowl depth at the rim is
\begin{equation}
  d_{\mathrm{bowl}} = R - \sqrt{R^2 - r_{\mathrm{out}}^2}
  \approx \SI{14.6}{\milli\meter} \approx 9.5\lambda,
\end{equation}
which is too deep for a single-plane time-delay projection to
capture the converging wavefront accurately.

\subsection{Plane-by-plane source injection}
\label{sec:bowl:injection}

We decompose the bowl surface into thin axial slices of thickness
$\Delta z$ aligned with the propagation step.  Each slice contains the
annular ring of transducer surface lying between $z_i$ and
$z_i + \Delta z$, where $z = 0$ is the bowl apex and
$z(r) = R - \sqrt{R^2 - r^2}$ maps lateral position to bowl depth.

For a given electronic focal depth $z_f$ (which may differ from the
geometric focus at $R$), the focusing delay at each surface point
$(x, y, z_{\mathrm{bowl}})$ is computed from the distance to the
focal point as
\begin{equation}
  d(x,y) = \sqrt{x^2 + y^2 + (z_f - z_{\mathrm{bowl}})^2},
  \qquad
  \tau(x,y) = \frac{d_{\max} - d(x,y)}{c_0},
  \label{eq:bowl_delay}
\end{equation}
so that the outermost elements fire first and all wavefronts converge
at $z_f$.  For an annular phased array, the per-pixel delays are
averaged within each element to produce per-element delays, matching
the hardware constraint.

During propagation, the solver starts from the apex slice
($z = 0$) and marches forward.  As it crosses each slice depth $z_i$,
the corresponding source field is \emph{added} to the propagating
wavefield,
\begin{equation}
  p(\mathbf{x}, t)\big|_{z = z_i^+}
  = p(\mathbf{x}, t)\big|_{z = z_i^-}
  + s_i(\mathbf{x}, t),
\end{equation}
where $s_i$ is the time-delayed pulse restricted to the annular ring
in the $i$-th slice, with its delay corrected by $-z_i/c_0$ to account
for the propagation distance already traversed from the apex.  This
injection is implemented identically to the phase-screen application,
as a conditional addition at the appropriate $z$ coordinate within
the main propagation loop.

\subsection{Fullwave~2 comparison}
\label{sec:bowl:results}

To validate the plane-by-plane model, we compare the ASM prediction
against a 3-D Fullwave~2 simulation in water.  The TIPS bowl is focused
electronically at \SI{50}{\milli\meter} (inside the geometric focus at
$R = \SI{80}{\milli\meter}$) using a \SI{1}{\mega\hertz}, 3-cycle
Gaussian-windowed pulse at $p_0 = \SI{400}{\kilo\pascal}$.

The Fullwave~2 reference uses the volumetric bowl source (3 voxel layers
along the surface normal) solved on a $623 \times 873 \times 873$ grid
at $\Delta x = \SI{0.128}{\milli\meter}$ ($\lambda/12$) with 6234 time
steps.  The ASM uses $\Delta x = \lambda/5$ with a
$365 \times 365 \times 425$ grid and 260 propagation steps
($\Delta z = \Delta x$).

\begin{figure}[htbp]
  \centering
  \includegraphics[width=\columnwidth]{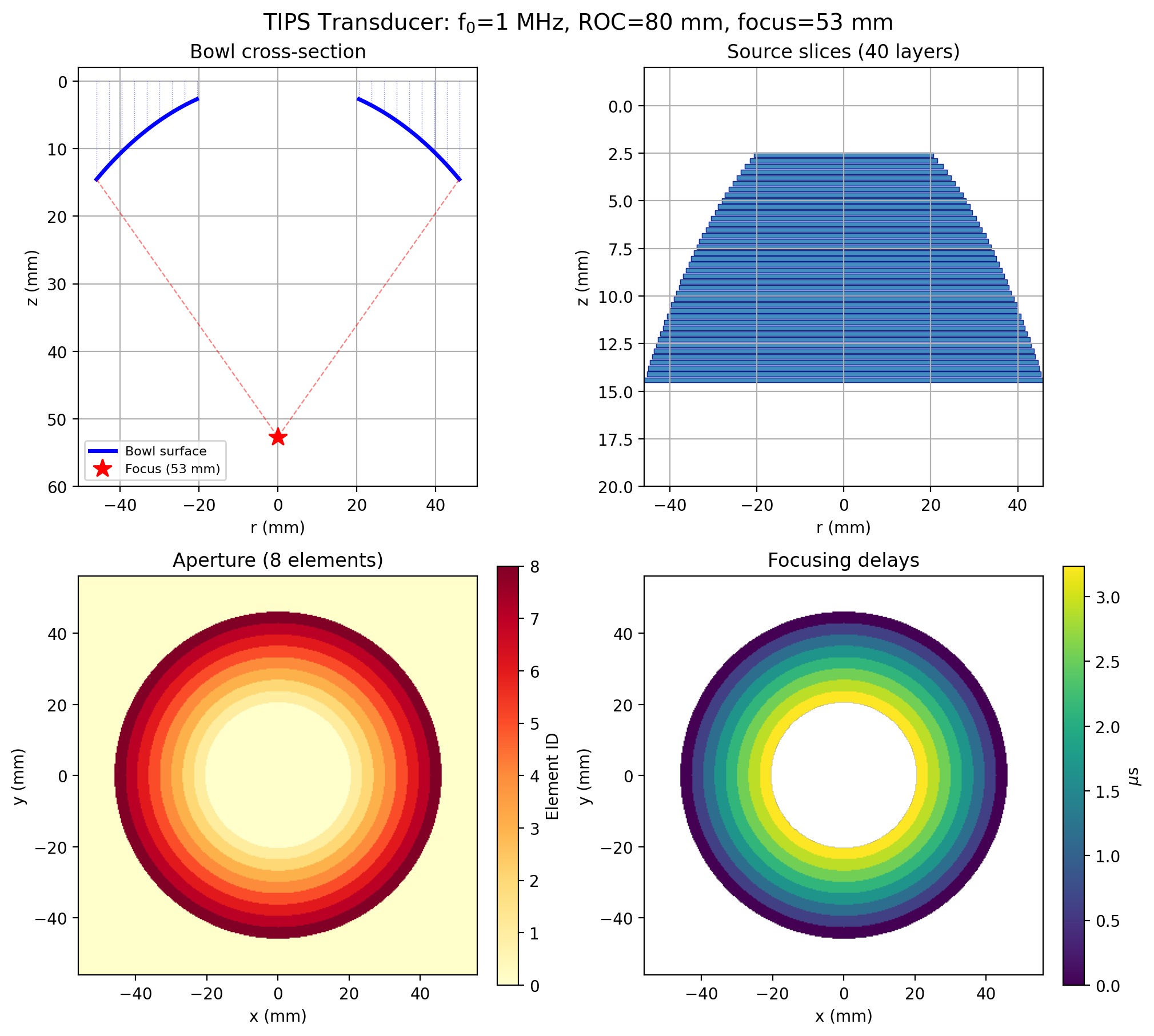}
  \caption{TIPS bowl transducer setup.  The panels show the bowl
    cross-section ($R = \SI{80}{\milli\meter}$) with element
    boundaries and focal lines (top left), the plane-by-plane source
    slices spanning the \SI{14.6}{\milli\meter} bowl depth (top
    right), the 8-element annular aperture (bottom left), and the
    per-element focusing delays (bottom right).}
  \label{fig:tips_setup}
\end{figure}

Figure~\ref{fig:tips_setup} shows the TIPS transducer geometry,
including the bowl cross-section with element boundaries and focal
lines, the plane-by-plane source slices spanning the
\SI{14.6}{\milli\meter} bowl depth, the 8-element annular aperture,
and the per-element focusing delays.

\begin{figure}[htbp]
  \centering
  \includegraphics[width=0.92\columnwidth]{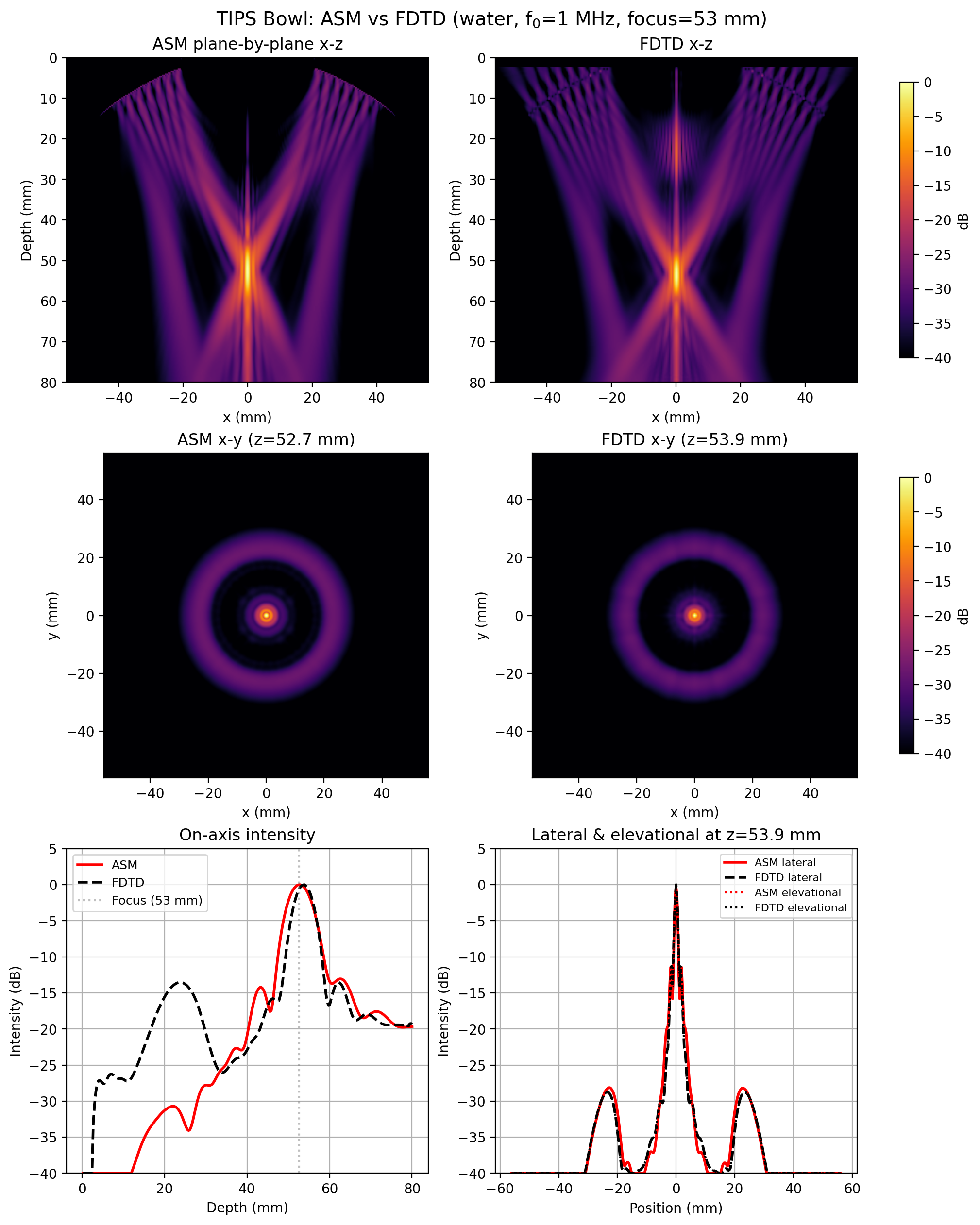}
  \caption{Comparison of the plane-by-plane ASM and Fullwave~2 for
    the TIPS bowl in water.  The top row shows axial ($x$--$z$) beam
    profiles and the middle row shows focal-plane ($x$--$y$)
    intensity maps at each method's focal depth.  The bottom row
    shows the on-axis intensity (left) and the lateral and
    elevational profiles at the Fullwave~2 focal depth (right).
    Each method is normalized to its own peak so the beam shapes can
    be compared directly.  The ASM focal depth
    (\SI{52.7}{\milli\meter}) matches the Fullwave~2 value
    (\SI{53.9}{\milli\meter}) to within 2.3\%.}
  \label{fig:tips_comparison}
\end{figure}

Figure~\ref{fig:tips_comparison} compares the plane-by-plane ASM
against the Fullwave~2 reference.  Each method is normalized to its own
peak intensity so the beam shapes can be compared directly.
The ASM focuses at $z = \SI{52.7}{\milli\meter}$, within 2.3\% of
the Fullwave~2 focal depth at \SI{53.9}{\milli\meter}.
The on-axis, lateral, and elevational beam profiles match well in
shape. The remaining differences are concentrated in the pre-focal
near-field where the annular aperture geometry produces
resolution-dependent interference fringes.
The ASM source amplitude is scaled by a factor of 1.31 to align
with the Fullwave~2 source convention.  After this scaling, both methods
produce a focal peak pressure of \SI{13.8}{\mega\pascal}.

The ASM requires substantially less memory than Fullwave~2.  The Fullwave~2
solver allocates the full 3-D volume
($623 \times 873 \times 873 \approx \SI{475}{M}$ grid points) for
approximately eight field arrays (pressure, velocity components,
material maps), totaling roughly \SI{15}{\giga\byte}.  The ASM
stores only the current 2-D-plus-time field
($365 \times 365 \times 425 \approx \SI{57}{M}$ points) together with
precomputed frequency-domain operators of the same shape, for a total
of approximately \SI{1.6}{\giga\byte}, a factor of nine reduction.
For the present benchmark Fullwave~2 completes in \SI{1477}{\second}
(multi-GPU) while the ASM requires \SI{517}{\second} on a single CPU,
a $2.9\times$ speed-up despite running without GPU acceleration.  The
ASM wall time is dominated by its 260 sequential FFT-based propagation
steps and could be reduced further with a coarser axial step.

% ------------------------------------------------------------------
\section{Discussion}
\label{sec:discussion}

The results show that the angular spectrum method, extended with
consistent obliquity corrections on the linear and nonlinear
operators, a shock-capturing flux-conservative nonlinear update, and
a plane-by-plane injection scheme for deeply curved sources,
reproduces focal fields of Fullwave~2 quality at a fraction of the
computational cost for transcranial and therapeutic
configurations.

The diffraction benchmarks confirm that the propagator correctly
handles unfocused and geometrically focused sources, matching
analytical Rayleigh--Fresnel and Airy solutions with RMS errors of
0.086 on axis and 0.002 at the focal plane.  The attenuation filter
reproduces the prescribed power-law absorption together with its
Kramers--Kronig dispersion, and the per-mode $k/k_z$ path-length
correction recovers the predicted $|\hat A|^{1/\cos\theta}$ scaling
on representative oblique plane-wave components.  The
Kurganov--Tadmor flux composed with Strang splitting achieves
second-order convergence in the strongly nonlinear regime, compared
with first-order for the sequential and Rusanov-based schemes.  The
KT flux resolves shocks more sharply than the first-order Rusanov
flux ($L^2$ error 0.036 versus 0.044), and the adaptive CFL
sub-cycling maintains stability at all tested step sizes without
loss of accuracy.

The three absorbing-boundary improvements act cumulatively, reducing
the periodic-FFT reflection ratio by a factor of 2.4 without
distorting the interior field.  The frequency-weighted spatial
damping preferentially attenuates low-frequency components that see
fewer wavelengths in the boundary layer.  The first-order
Engquist--Majda super-absorbing correction selectively removes
incoming wave energy, and the Wendland $C^2$ taper eliminates the
impedance discontinuity that a plain quadratic profile introduces at
the onset of damping.

In the transcranial benchmark the phase-and-amplitude-screen ASM
matches the Fullwave~2 focal-plane intensity with 1.1\% RMS difference and
predicts a through-skull insertion loss of \SI{5.4}{\deci\bel},
consistent with measured values of
\SIrange{5}{10}{\deci\bel}~\cite{pinton2012}.  The focal depth
differs by only 2.6\% from the Fullwave~2 reference.  The TIPS bowl
validation extends this agreement to a curved-source geometry.  The
plane-by-plane ASM matches the Fullwave~2 focal depth to within 2.3\%
with $9\times$ less memory and $2.9\times$ less wall
time.  The rim-ray obliquity at the aperture edge of this bowl
approaches $30^\circ$, a regime in which the per-mode attenuation
correction and the beam-averaged nonlinear correction are both
numerically active, although a controlled ablation study separating
their individual contributions is left to future work.

Several limitations remain.  The ASM is a forward-only propagator.
The phase-and-amplitude-screen model captures the dominant wavefront
aberration and insertion loss but neglects reflection, mode
conversion, and multiple scattering at acoustic interfaces.  The
remaining discrepancy with Fullwave~2 in the transcranial benchmark is
concentrated in the pre-focal region and near the skull surface
where these backward-propagating effects are strongest.  The
obliquity correction on the nonlinear operator is applied as a
power-weighted scalar rather than per plane-wave mode.  This
approximation is faithful when a single propagation direction
dominates at each axial station, but it loses accuracy where
plane-wave fronts of comparable amplitude cross, such as near the
geometric focus.  The super-absorbing
boundary uses a first-order Engquist--Majda decomposition that is
exact only for normally incident waves, and a higher-order
formulation \cite{higdon1986} could improve performance for
wide-angle beams.

Beyond the longitudinal-acoustic case, the same flux-conservative
nonlinear update is applicable to the cubic Burgers equation that
governs soft-tissue shear shocks
\cite{catheline2003,pinton2010,giammarinaro2016}.  A sonic-detector
patch on the Kurganov--Tadmor flux suppresses spurious dissipation
at sub-cell sonic crossings of the non-convex flux $f(u)=-u^{3}/3$.
With this patch the scheme converges to the analytical entropy
solution on canonical Riemann benchmarks covering shock,
rarefaction, and compound-wave behavior (\Cref{app:cubic}), with
$L^{2}$ errors of $6\!\times\!10^{-3}$, $5\!\times\!10^{-4}$, and
$3\!\times\!10^{-2}$ at $z=1$, respectively.

% ------------------------------------------------------------------
\section{Conclusion}
\label{sec:conclusion}

Three developments extend the angular spectrum method to strongly
nonlinear transcranial and therapeutic propagation from deeply
curved bowl transducers.  First, the split-step update carries a
consistent obliquity correction in both the linear and nonlinear
operators.  A per-mode $k/k_z$ factor in the attenuation and
dispersion filter restores the correct path length
$\Delta z/\cos\theta$ for every plane-wave component, and a
beam-averaged scaling of the Burgers coefficient by
$\langle k/k_z \rangle$ tracks the beam's actual angular content in
the nonlinear step.  Second, the retarded-time Burgers operator is
discretized with a second-order Kurganov--Tadmor central-upwind flux
using MUSCL reconstruction and SSP-RK2 time integration, composed
with the diffraction and attenuation operators through Strang
splitting with adaptive CFL sub-cycling.  This discretization
resolves fully developed shocks without the temporal refinement
imposed by the CFL coupling intrinsic to FDTD.  Third, a
plane-by-plane injection scheme decomposes deeply curved bowl
transducers into axial slices that are injected at their correct
propagation depths, preserving the aperture-dependent
shock-formation distance that governs near-focal nonlinear
accumulation.

Phase-and-amplitude screens derived from CT data incorporate
skull-induced aberration and frequency-dependent insertion loss into
the FFT-based propagator.  In a transcranial benchmark the ASM
matches a Fullwave~2 reference to 1.1\% RMS in focal-plane intensity and
predicts a through-skull insertion loss of \SI{5.4}{\deci\bel},
consistent with measured values \cite{pinton2012}.  For a clinical
TIPS annular bowl array, the method matches the Fullwave~2 focal depth to
within 2.3\% with $9\times$ less memory and $2.9\times$
less wall time.  Three enhanced absorbing-boundary treatments reduce
periodic-FFT reflections by a factor of 2.4 without distorting the
interior field.  A spatially resolved intensity-loss field,
accounting for the full harmonic content of the nonlinear wave,
provides an absorbed-energy map suitable for acoustic radiation
force and thermal dose calculations.

These developments extend the
angular spectrum method from homogeneous free-field propagation to
heterogeneous transcranial paths and immersed curved-source
geometries, at computational costs compatible with treatment
planning and aberration-correction workflows.  The solver can also
serve as the acoustic front end of end-to-end modelling chains such
as the companion open-source ultrasound-neuromodulation framework
\cite{pinton2026neuromod}.

\funding{This work was supported by the National Institutes of Health
under grants R01-EB037345 and R01-EB036295.}

\data{The source code and all validation scripts needed to reproduce
the figures in this paper are openly available under the Apache~2.0
license at
\url{https://github.com/pinton-lab/angularspectrum_heterogeneous}.}

\bibliographystyle{IEEEtran}
\bibliography{references}

\appendix

\section{Cubic Burgers extension for the shear-shock regime}
\label{app:cubic}

Soft-tissue shear waves are dominated by cubic rather than
quadratic nonlinearity \cite{catheline2003,gennisson2007}, and the
resulting shear-shock dynamics, including reflection at tissue
interfaces, have been characterized theoretically and numerically
in earlier work \cite{pinton2010,giammarinaro2016}, and a reduced
viscoelastic FDTD formulation for the same ultrasound-driven shear-wave
regime is developed in a companion paper \cite{pinton2026shear}.  At the
1--10\,m/s shear speed typical of brain tissue, the cubic
nonlinear coefficient $\beta_{3}\!\sim\!10^{2}$--$10^{3}$ drives
shock formation in less than one wavelength even at modest
particle velocities, and the absorption length is only a few
wavelengths.  This regime is qualitatively different from the
longitudinal-acoustic case treated in the body of the paper, but
its propagation operator embeds cleanly into the same
flux-conservative angular-spectrum framework.  Only the nonlinear
sub-step is replaced.

\subsection{Equation form and discretization}

The cubic Burgers equation in retarded time reads
\begin{equation}
\frac{\partial p}{\partial z} \;=\; -\,N_{3}\,
\frac{\partial}{\partial \tau}\!\left(p^{3}\right),
\qquad
N_{3}\!=\!\frac{\beta_{3}}{3\,c_{0}^{5}\,\rho_{0}^{2}},
\label{eq:cubic-burgers}
\end{equation}
with characteristic speed $\lambda(p)=-p^{2}$, which is always
non-positive and sonic at $p=0$.  In contrast to quadratic Burgers,
the conservation flux $f(p)=-p^{3}/3$ is non-convex.  The second
derivative $f''=-2p$ changes sign at $p=0$, so the equation admits
compound waves consisting of rarefaction-shock combinations at
sign-changing states.  The same coefficient $N_{3}$ enters the
Rusanov and Kurganov--Tadmor (KT) march variants without
modification.  The schemes differ only in their flux discretization.
Standard Rusanov uses the symmetric flux
$F = -(u_{L}^{3}+u_{R}^{3})/3 - \lambda_{1/2}(u_{R}-u_{L})$,
whereas KT reduces to $f_{\pm} = -u_{\pm}^{3}/3$ in the smooth
limit.

The cubic flux kernels are direct analogues of the existing
quadratic ones, with $f(u)=-u^{2}/2$ replaced by $-u^{3}/3$ and
$\lambda(u)=|u|$ replaced by $u^{2}$.  Each march-step variant
(sequential, split-step standard, TVD-Rusanov, and Strang-split
second-order KT, with and without nonlinearity-obliquity correction
and per-pixel attenuation-loss tracking) has a one-line cubic
counterpart that substitutes the inner flux call.  The CFL margin
uses $|p|^{2}$ rather than $|p|$ so that the adaptive step-size
control retains the correct amplitude dependence in the cubic regime.
Selection between the quadratic and cubic operators is made at
solver configuration, with the quadratic regime as the default.

\subsection{Riemann-problem validation}
\label{app:cubic-riemann}

The cubic flux kernels are validated against analytical entropy
solutions of three one-dimensional Riemann problems that cover
compressive shock, rarefaction, and non-convex compound-wave
behavior.  Each test runs on a single-pixel transverse column
($1\times 1\times n_{T}$) so that only the nonlinear sub-step
operates, which isolates the flux discretization from diffraction
and attenuation.

The first problem (Test~A) is a compressive shock with $u_{L}=0$
and $u_{R}=+1$.  The Rankine--Hugoniot condition gives the shock
speed $s=-(u_{L}^{2}+u_{L}u_{R}+u_{R}^{2})/3=-1/3$, which is
Lax-admissible.  The exact solution at depth $z$ is a step at
$\tau = sz$.

The second problem (Test~B) is a rarefaction with $u_{L}=+1$ and
$u_{R}=0$.  The exact solution is the self-similar fan
$u(\tau,z) = \sqrt{-\tau/z}$ for $-z\!\le\!\tau\!\le\!0$, with
$u=u_{L}$ for $\tau<-z$ and $u=u_{R}$ for $\tau>0$.

The third problem (Test~C) is a compound wave with $u_{L}=+1$ and
$u_{R}=-1$.  The Oleinik upper-concave-envelope construction on
$[u_{R}, u_{L}]$ predicts a rarefaction from $u_{L}=1$ to an
intermediate state $u^{\star}=1/2$, where the envelope touches the
flux curve at the tangent point with slope $-1/4$, followed by a
sonic shock from $u^{\star}=1/2$ to $u_{R}=-1$ at speed $-1/4$.
This compound structure is characteristic of non-convex flux and
probes the scheme's ability to resolve mixed wave types.

\Cref{tab:cubic-riemann-errors} summarizes $L^{2}$ errors against
the analytical solutions on a 4001-cell grid at $z=1$ (CFL$\approx 0.4$
for $|u|\le 1$).

\begin{table}[htbp]
  \centering
  \begin{tabular}{lccc}
    \hline
    Scheme & Test A (shock) & Test B (rarefaction) & Test C (compound) \\
    \hline
    Rusanov     & $0.008$ & $0.0018$ & $0.012$  \\
    Rusanov-TVD$^{\dagger}$ & $0.157$ & $0.206$  & $0.518$ \\
    KT$^{\ast}$ & $0.006$ & $0.00046$ & $0.026$  \\
    \hline
  \end{tabular}
  \caption{$L^{2}$ errors of the cubic-Burgers flux kernels against
    analytical entropy solutions for three Riemann problems at $z=1$.
    $^{\dagger}$The minmod-style limiter over-damps both shocks
    (Test~A) and rarefactions (Test~B), and on non-convex flux it
    selects the inadmissible single-shock branch through the sonic
    point (Test~C, where the compound-wave structure is missed
    entirely).  This is a known limitation of TVD limiters on
    non-convex flux.  $^{\ast}$KT uses the sonic-detector switch
    described in \Cref{app:kt-sonic-fix}.  Faces whose
    MUSCL-reconstructed states or cell-center states straddle
    $u=0$, together with their two immediate neighbors, fall back
    to a cell-center Rusanov flux.  This substitution suppresses
    the rarefaction over-prediction near the sonic-shock junction
    observed in Test~C without sacrificing second-order accuracy on
    smooth regions, and the errors on Tests~A and~B remain at the
    KT level.  With the fix in place both KT and standard Rusanov
    track the entropy solution accurately in all three cases.  KT
    remains uniformly second-order on smooth data and is the
    recommended choice for the body's quadratic-Burgers physics,
    which has no sonic points.}
  \label{tab:cubic-riemann-errors}
\end{table}

\begin{figure}[htbp]
  \centering
  \includegraphics[width=0.9\columnwidth]{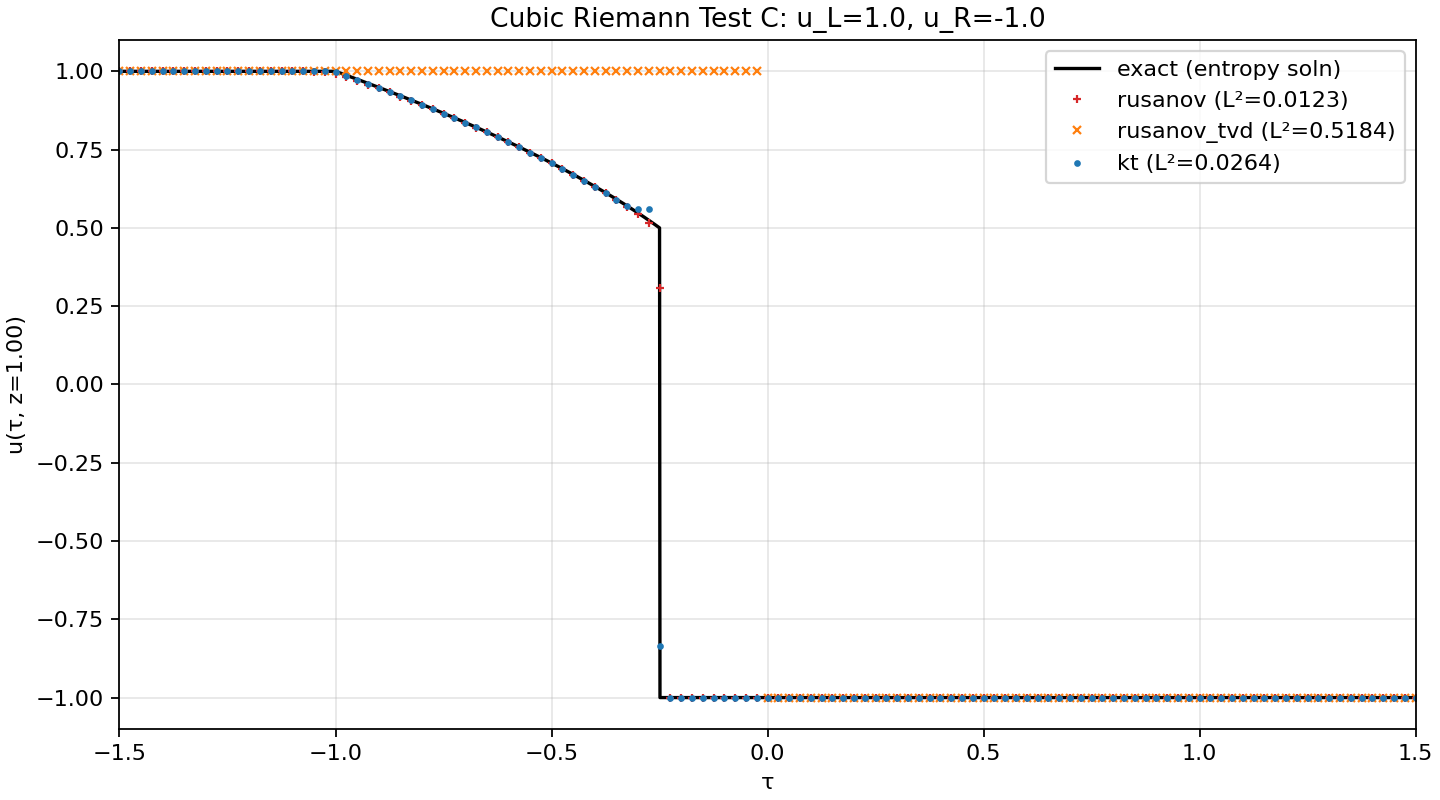}
  \caption{Cubic-Burgers Riemann Test~C ($u_{L}=+1$, $u_{R}=-1$) at
    $z=1$.  The solid line shows the exact entropy solution, with
    the rarefaction $u=\sqrt{-\tau/z}$ on $[-z, -z/4]$ and the sonic
    shock at $\tau=-z/4$.  Markers show standard Rusanov (red $+$),
    TVD-Rusanov (orange $\times$), and KT (blue dots).  The compound
    wave is the canonical stress test for non-convex flux schemes.}
  \label{fig:cubic-riemann-C}
\end{figure}

\Cref{fig:cubic-riemann-C} shows the most discriminating case
(Test~C). The Rusanov and KT schemes faithfully capture both the
$+\sqrt{-\tau/z}$ rarefaction and the sonic shock at $\tau=-z/4$,
while TVD-Rusanov misses the rarefaction entirely.

\subsection{Sonic-detector fix in the KT cubic flux}
\label{app:kt-sonic-fix}

The Kurganov--Tadmor central-upwind flux is built around one-sided
local wave-speed estimates $a^{\pm}=\max/\min(0,f'(u^{L}),f'(u^{R}))$
and the denominator $a^{+}\!-\!a^{-}$.  For the cubic flux
$f(u)=-u^{3}/3$ with $f'(u)=-u^{2}\!\le\!0$, both $a^{\pm}$ are
non-positive, $a^{+}\!=\!0$ wherever $\max(|u^{L}|,|u^{R}|)\!>\!0$,
and $a^{+}\!-\!a^{-}\!=\!\max((u^{L})^{2},(u^{R})^{2})$.  In smooth
regions this formula yields a one-sided upwind from the right state
with zero numerical viscosity, which accounts for the second-order
behavior of KT on the pure rarefaction (Test~B).  Near the sonic
point $u\!=\!0$, however, two distinct failure modes arise.

The first failure mode is a denominator collapse.  As both states
approach $u\!=\!0$, the difference $a^{+}\!-\!a^{-}\!\to 0$ and the
central-upwind formula degenerates.  A Lax--Friedrichs fallback is
then required for numerical stability.

The second failure mode occurs even where the denominator is
well-conditioned.  When the MUSCL-reconstructed cell-edge states
straddle $u\!=\!0$, the sub-cell linear reconstruction implies a
sonic crossing inside the cell that the central-upwind formula
does not resolve.  This manifests as a systematic positive bias in
the rarefaction region of Test~C, where the unpatched KT result
lies visibly above the analytical $\sqrt{-\tau/z}$ curve
(\Cref{fig:cubic-riemann-C}).

Both failure modes are addressed by detecting sonic faces and
reverting their flux to a cell-center Rusanov form on those faces
only,
\begin{equation}
F_{j+1/2}^{\text{Rusanov}} \;=\;
\tfrac{1}{2}\!\left[-\tfrac{u_{j}^{3}+u_{j+1}^{3}}{3}\right]
- \tfrac{1}{2}\,\max(u_{j}^{2},u_{j+1}^{2})\,(u_{j+1}-u_{j}),
\label{eq:kt-sonic-rusanov}
\end{equation}
which has the correct entropy behavior through the sonic point
without the Lax--Friedrichs over-dissipation.  A face is flagged
sonic when any of three conditions holds, namely (a)
$u^{L}u^{R}\!<\!0$ on the MUSCL-reconstructed states, (b)
$u_{j}u_{j+1}\!<\!0$ on the cell-center states, or (c)
$|a^{+}\!-\!a^{-}|\!<\!10^{-14}$.  The flag is then dilated by one
face to either side, so that a sonic crossing's two immediate
face neighbours also fall back to \Cref{eq:kt-sonic-rusanov}.
This dilation reduces the KT Test~C error from
$7.6\!\times\!10^{-2}$ to $2.6\!\times\!10^{-2}$.  Inspection of
the residual error budget confirms that the rarefaction
over-prediction is caused by MUSCL artifacts on the adjacent
faces, which the dilated mask captures, rather than by the sonic
face itself.

The penalty on smooth data is modest.  For Test~B, which has no
sonic crossing of any kind, the error rises only marginally from
$3.5\!\times\!10^{-4}$ to $4.6\!\times\!10^{-4}$.  The dilation
triggers there only at the rarefaction's right edge where
$u\!\to\!0$ on the upper side, so that a few cells lose their
second-order treatment.  Test~A, in which $u\!\in\![0,1]$
everywhere except in the post-shock state, is unaffected.  Across
all three Riemann tests the patched KT scheme is uniformly best on
smooth data and remains within a factor of $2.2$ of plain Rusanov
on the compound-wave stress test, and it preserves second-order
convergence in the quadratic-Burgers regime used in the body of
the paper.

\end{document}